\documentclass[14pt,twocolumn,superscriptaddress]{revtex4}

\usepackage{amsmath,psfrag,times,graphicx}

\def\U#1{{%
\def\O{\mbox{O}}
\def\u{\mbox{u}}
\mathcode`\u=\mu
\mathcode`\O=\Omega
\mathrm{#1}}}
\def\sub#1{_{\scriptsize\mbox{#1}}}

\def\fracpd#1#2{\frac{\partial#1}{\partial#2}}
\def\E{\mathcal E}
\def\P{\mathcal P}
\def\cc{\mbox{c.c.}}
\def\intinf{\int^{\infty}_{-\infty}}

\begin{document}

\title{\vspace*{4mm}Simulation of Slow Light with Electronics Circuits}
\author{T. Nakanishi}
\author{K. Sugiyama}
\author{M. Kitano}
\affiliation{Department of Electronic Science and Engineering,
Kyoto University, Kyoto 615-8510, Japan}
\affiliation{CREST, Japan Science and Technology Agency, Tokyo 103-0028, Japan}
\date{\today}

\begin{abstract}
\vspace*{4mm}
We present an electronic circuit which simulates wave propagation
in dispersive media.
The circuit is an array of phase shifter composed of
operational amplifiers and can be described with
a discretized version of one-dimensional wave equation for
envelopes.
The group velocity can be changed both spatially and
temporarily.
It is used to emulate slow light or stopped light, which has been realized
in a medium with electromagnetically induced transparency (EIT).
The group-velocity control of optical pulses is expected to be a
useful tool in the field of quantum information and communication.
\vspace*{5mm}
\end{abstract}

\maketitle

\section{Introduction\label{sec:intro}}

In an optical medium, the envelope of a light pulse travels
at a group velocity $v\sub{g}$,
which is defined as the reciprocal slope of 
the dispersion relation for the medium.
Recent advances in controlling the dispersive properties
have enabled superluminal or negative group velocities \cite{Wang,Dogariu},
extremely slow group velocities \cite{Hau},
and even stopping light pulses \cite{Phillips,Liu}.

In a classical textbook \cite{Brillouin},
Brillouin pointed out that
the group velocity can exceed the light speed $c$ in a vacuum
and such a superluminal phenomenon does not conflict with
causality.
However, even after several experimental demonstrations \cite{Wang,Dogariu},
the superluminal light propagation has still invited some misunderstandings,
because it is very counterintuitive that 
a light pulse runs faster than $c$.
Recently we have proposed an electronic circuit
which exhibits negative group delays.
It produces an output pulse ahead of 
the input pulse \cite{Nakanishi,Kitano}. 
This circuit shares the same principle as 
light propagation at superluminal or negative group velocities.
The circuit is so simple
that one can easily build it and  
examine how this puzzling phenomenon takes place.

Extremely slow light has been realized
by utilizing electromagnetically induced transparency
(EIT) \cite{Harris}.
In the EIT experiments,
probe light and control light are tuned to a common excited level of
$\Lambda$-type three-level atoms.
The two light beams destructively interfere with each other 
through the common upper level
and the light absorption disappears.
The refractive index for the probe light,
as well as the absorption coefficient,
varies largely within very narrow bandwidth,
called a ``transparency window,'' where the absorption is vanished.
Hence in the presence of the control light,
the group velocity of the probe light becomes extremely small,
compared with $c$.
Hau {\it et al.}  \cite{Hau} succeeded in reducing the group velocity
down to $17\,\U{m/s}$ in laser-cooled atomic gas.
They prepared a single-shot pulse with the duration of $2.5\,\U{us}$
and the pulse length of $750\,\U{m}$ in a vacuum,
and fed it into the EIT medium.
Owing to the slow down,
the pulse length was compressed into $43\,\U{um}$ in the medium.

The group velocity is proportional to
the width of the transparency window, which is determined
by the intensity of the control light;
the weaker the control light, the smaller the group velocity.
However, we cannot slow down the probe pulse unconditionally.
The lower limit of the group velocity is given
by the condition that the transparency window 
should be wider than the bandwidth of the probe pulse.
Otherwise the pulse shape would be deformed by the absorption.
This condition is often called the EIT condition.

This limitation can be overcome by changing the intensity of
the control light when the probe pulse is traveling in
the EIT medium \cite{Fleischhauer1,Fleischhauer2}.
The intensity of the control light is kept constant
until the entire probe pulse fits into the EIT medium.
Subsequently the control light is turned down smoothly,
and the probe pulse is slowed down with its spatial profile unchanged.
The pulse is stopped in the medium when the control light is
eventually turned off.
During the stopping process,
the spectral width of the probe pulse is compressed 
and can be fitted in the transparency window,
which narrows down according to the dimming of control light.
The stopped pulse can be restarted
by restoring the control light.
The released pulse is identical with the original one,
because all the processes, freezing and releasing, are carried out
coherently.
The characteristics of light pulses, such as the amplitude, phase, and 
beam profile, can be stored.
Even the quantum state of light could be stored in the medium,
and this method is promising as a quantum memory.

This paper deals with the physics of the slow and stopped light,
and introduces a circuit which simulates the light propagation
with velocity control.
This circuit model is very clear and instructive
as that for the superluminal propagation.
Moreover it is easy to construct the circuits and change the
circuit parameters freely.
We can set the time constant slow enough that
the propagation can be seen as blinks of
a series of light emitting diodes (LEDs), which monitor
the output of each stage.
The circuit behaves like a running message display of single row 
but the operation principle is quite different. 

In Sec.~\ref{Sec:theory} we give a theoretical framework of pulse 
propagation in dispersive media
and introduce the circuit for the simulation.
In Sec.~\ref{Sec:experiment} we show several experimental results,
which correspond to slow light or stopped light experiments,
and summary is given in Sec.~\ref{Sec:discuss}.

\section{Theory\label{Sec:theory}}
\subsection{Pulse propagation and group velocity}

In a medium, 
the wave equation in one dimension with respect to 
the electric field, $E(x, t)$, is given by
\begin{align}
 \fracpd{^2 E}{^2 t} - c^2 \fracpd{^2 E}{^2 x} = -
 \frac{1}{\epsilon_0} \fracpd{^2 P}{^2 t},
 \label{Eq:wave}
\end{align}
where $P(x, t)$ is the polarization of the medium induced by the electric field.
We introduce the envelopes, $\E(x, t)$ and $\P(x, t)$ as
\begin{align}
 E(x, t) = \E(x, t) \, e^{i (\omega_0 t - k x)} + \cc, \\
 P(x, t) = \P(x, t) \, e^{i (\omega_0 t - k x)} + \cc,
\end{align}
where $\cc$ stands for the complex conjugate of the preceding term.
The carrier frequency $\omega_0$ is located near the center of the
spectrum of the electric field.
The associated wavenumber $k$ will be determined later on.

Assuming that $\E(x, t)$ and $\P(x,t)$ are slowly varying functions
with respect to space and time, 
we can reduce the order of differentiations of
each term of Eq.~(\ref{Eq:wave}) in the following way:
\begin{align}
 \fracpd{^2E}{t^2}
 &\simeq \left[
 2 i \omega_0 \fracpd{\E}{t} - \omega_0^2 \E
 \right]  \, e^{i (\omega_0 t - k x)}  + \cc, \label{Eq:1} \\
 c^2 \fracpd{^2 E}{x^2} &\simeq c^2 \left[
 - 2 i k \fracpd{\E}{x} - k^2 \E
 \right]  \, e^{i (\omega_0 t - k x)} + \cc,   \label{Eq:2}\\
 \frac{1}{\epsilon_0} \fracpd{^2 P}{t^2} &\simeq
 \frac{1}{\epsilon_0} \left[
 2 i \omega_0 \fracpd{\P}{t} - \omega_0^2 \P
 \right] \, e^{i (\omega_0 t - k x)}  + \cc \label{Eq:3a} 
\end{align}
For slowly varying envelopes, 
the spectrum of the electric field is confined within
the narrow regions: 
$[\pm\omega_0-\delta\omega/2, \pm\omega_0+\delta\omega/2]$,
where $\delta\omega$ represents the spectral width.
We have assumed that fractional changes of envelopes
over a period or a wavelength are negligibly small. 

Then the susceptibility of the medium, $\chi(\omega, x)$,
can be regarded as a linear function of $\omega$ within the bandwidth
and the relation between $\E(t, x)$ and $\P(t ,x)$
can be written as
\begin{align}
 \P (x, t) = \epsilon_0 \chi_0 \E(t, x) - i \epsilon_0 \chi_1 
 \fracpd{\E}{t} (x, t), \label{Eq:polarization}
\end{align}
where $\chi_0 = \chi(\omega_0),
 \chi_1 = {d \chi}/{d \omega}(\omega_0)$.
The derivation of Eq.~(\ref{Eq:polarization}) 
is given in Appendix \ref{appendix}.
With the help of Eq.~(\ref{Eq:polarization}),
Eq.~(\ref{Eq:3a}) can be represented as
\begin{align}
 \frac{1}{\epsilon_0} \fracpd{^2 P}{t^2} \simeq 
 \left[i \omega_0 ( 2 \chi_0 + \omega_0 \chi_1) \fracpd{\E}{t}
 - \omega_0^2 \chi_0 \E \right] + \cc  \label{Eq:3b}
\end{align}
By substituting Eqs.~(\ref{Eq:1}), (\ref{Eq:2}), and (\ref{Eq:3b})
into the wave equation (\ref{Eq:wave}),
we obtain
\begin{align}
 \left[
 2 i \omega_0 \left(1+\chi_0+\frac{\omega_0 \chi_1}{2}\right) 
 \fracpd{\E}{ t} + 2 i c^2 k \fracpd{\E}{x} 
 \right] \nonumber \\
 - \left[ \omega_0^2(1+\chi_0) - c^2 k^2 \right] \E = 0. \label{Eq:wave2}
\end{align}
Considering the case of monochromatic fields
($\partial \E/\partial t= \partial \E/\partial x=0$), 
we can determine the wavenumber 
$k = n \omega_0 / c$,
where $n \equiv \sqrt{1+\chi_0}$ is the refractive index.

Then the former terms of Eq.~(\ref{Eq:wave2}) yield
the wave equation for the envelope as
\begin{align}
 \fracpd{\E}{t} + v\sub{g} \fracpd{\E}{x} = 0,
 \label{Eq:envelope}
\end{align}
where
\begin{align}
 v\sub{g} \equiv \frac{c}{\displaystyle n+\frac{\omega_0 \chi_1}{2n}},
\label{Eq:groupvelocity}
\end{align}
is the group velocity.

\subsection{EIT condition}
For EIT media, $\chi(\omega)$ changes sharply within the
EIT window, whose width is $w$.
The slope of $\chi$ is estimated as
$\chi_1=d\chi/d\omega\sim \alpha/w$, where
$\alpha$ is the variation of $\chi$ in the window.
Normally $\alpha$ corresponds to the optical thickness of the medium
and can be of the order of unity.
Assuming $n\sim 1$,
we have an estimation
\begin{align}
v\sub{g}\sim\frac{2 w}{\omega_0}c 
,
\end{align}
for the group velocity (\ref{Eq:groupvelocity}).
Roughly speaking, $w$ is equal to the
Rabi frequency of the control light $\Omega\sub{c}$,
which is proportional to the amplitude of 
the control light \cite{Fleischhauer1}.
The width $w$ can be effectively reduced down to the
atomic decoherence rate, which can be $100\,\U{kHz}$ or even smaller.
Therefore the extremely slow velocity, $v\sub{g}\sim 10^{-9}c$,
or virtual stopping can be possible. 

The spectrum of the pulse must be contained within the EIT window.
Namely, the spectral width $\delta\omega$ of the envelope must satisfy
the condition:
\begin{align}
\delta\omega < w .
\end{align}
This linkage between the group velocity and the bandwidth condition
is a very important issue in slow light experiments.
A similar restriction comes into play in cases of
superluminal or negative group velocities \cite{Nakanishi,Kitano}.

\subsection{Solutions}

If $v\sub{g}$ is constant in time and space,
it is easy to show that 
the solution of Eq.~(\ref{Eq:envelope}) is $\E(t-x/v\sub{g})$.
This means that the envelope propagates in the medium 
at the group velocity $v\sub{g}$, keeping its shape.

When $v\sub{g}$ is a function of $x$,
the solution of  Eq.~(\ref{Eq:envelope}) can be written as
\begin{align}
 \E(t, x) = \phi \left(t- \int^{x}_0  
 \frac{d x^\prime}{v\sub{g}(x^\prime) }\right),
\end{align}
where $\E(t, 0) \equiv \phi(t)$ is the boundary condition at $x=0$.
The temporal profile of the pulse, or the spectrum,
is conserved at any $x$,
while the spatial profile shrinks (stretches)
at the location of small (large) $v\sub{g}(x)$.
This is the reason why the pulse length is compressed when
the light pulse is fed from the vacuum into the EIT medium.

Although the group velocity could be decreased further
by reducing the width $w$ of the transparency window,
$w$ must be wider than 
the bandwidth of the pulse, $\delta\omega$,
which is constant regardless of $v\sub{g}(x)$.
Therefore, the spectral width $\delta\omega$ of the
initial pulse sets the lower limit of the velocity.

On the other hand,
when $v\sub{g}$ is a function of $t$,
the solution of Eq.~(\ref{Eq:envelope}) is given by
\begin{align}
 \E(t, x) = \psi \left(x- \int^{t}_0  v\sub{g}(t^\prime) 
 d t^\prime \right),
\end{align}
where $\E(0, x) \equiv \psi(x)$ is the initial condition at $t=0$.
In this case,
the spatial profile of the pulse is identical at any time $t$,
but the temporal profile
is changed according to $v\sub{g}(t)$.
At time when $v\sub{g}(t)$ becomes small,
the temporal profile, i.e. the pulse width, spreads out,
and the frequency spectrum is compressed.

If the EIT condition is satisfied initially,
it is fulfilled all the time no matter how far we diminish 
the velocity $v\sub{g}$ or the transparency window $w$,
because the spectral width $\delta\omega$ is decreased
accordingly.
With this method, the stopping of light has been carried out.

\subsection{A circuit model}

\begin{figure}[h]
 \begin{center}
  \psfrag{n1}[c][c]{$v_n$}
  \psfrag{n2}[c][c]{$v_{n+1}$}
  \includegraphics[scale=1]{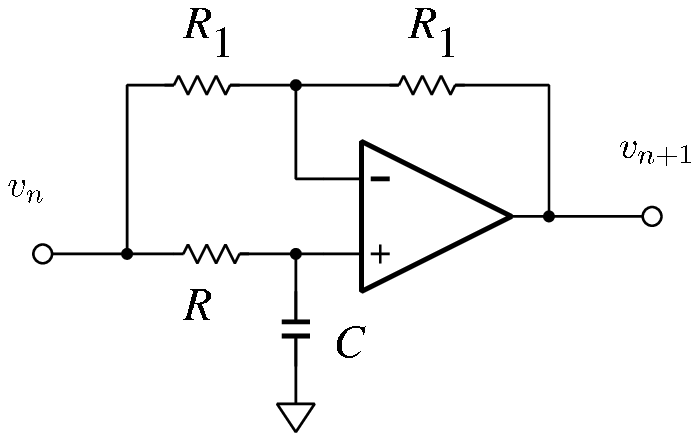}
  \caption{Elementary circuit for the simulation.}
  \label{Fig:allpass}
  \psfrag{v1}[c][c]{$v_1$}
  \psfrag{v2}[c][c]{$v_2$}
  \psfrag{v3}[c][c]{$v_3$}
  \psfrag{v4}[c][c]{$v_4$}
  \includegraphics[]{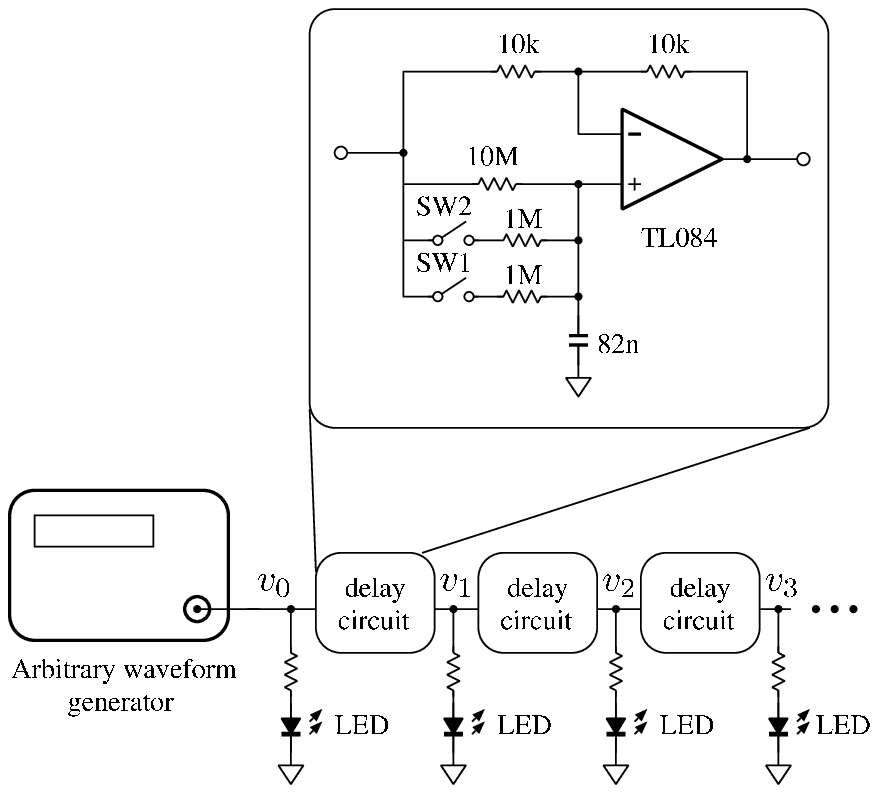}
  \caption{Schematic diagram of overall setup and circuit parameters.}
  \label{Fig:allpass-p}
 \end{center}
\end{figure}

We will show how to simulate the pulse propagation
described by Eq.~(\ref{Eq:envelope}) using
an electronic circuit.
First, 
we divide $x$ into uniform steps as $x=n \Delta x$ ($n$ is an integer).
We discretize the second term of Eq.~(\ref{Eq:envelope}) as
\begin{align}
 v\sub{g} \fracpd{\E}{x} \rightarrow
 \frac{1}{T}
 \{ v_{n+1}(t)
 -v_n(t) \},
 \label{Eq:disc}
\end{align}
where $T \equiv \Delta x / v\sub{g}$ and 
$v_n(t) \equiv \E(n \Delta x, t)$.
We approximate the first term of Eq.~(\ref{Eq:envelope})
as
\begin{align}
 \fracpd{\E}{t} \rightarrow
 \frac{d}{d t} \left\{
 \frac{v_{n+1}(t)+v_n(t)}{2}  \right\}.
 \label{Eq:app1}
\end{align}
Then the discretized version of Eq.~(\ref{Eq:envelope})
is obtained;
\begin{align}
  \frac{d}{d t} \left[
 \frac{v_{n+1}(t)+v_n(t)}{2}  \right] +
 \frac{1}{T}
[ v_{n+1}(t)
 -v_n(t) ] = 0.
 \label{Eq:discretize}			  
\end{align}
Introducing the Fourier transform of $v_n(t)$:
\begin{align}
 V_n(\omega) = \intinf v_n(t) \, e^{-i \omega t} \, dt,
\end{align}
we have the difference equation
\begin{align}
V_{n+1} (\omega) = H(\omega) V_n(\omega)
,
\end{align}
where 
\begin{align}
 H(\omega)  =
 \frac{1-i \omega T/2}{1+i\omega T/2}.
  \label{Eq:transfer}
\end{align}
is the transfer function.

The circuit illustrated in Fig.~\ref{Fig:allpass} has
the transfer function of Eq.~(\ref{Eq:transfer}).
It is well known as an all-pass filter,
which is used as a phase shifter \cite{Tietze}. 
The circuit provides a group delay $T$ for signals
whose spectrum is limited within the bandwidth $1/T$ \cite{Kitano}.
Replacing the medium with the length $\Delta x$
by the circuit with $T=\Delta x/v\sub{g}$
and cascading the circuits,
we can simulate the envelope propagation described by Eq.~(\ref{Eq:envelope}).

The quantity $\nu\sub{d} \equiv 1/T$, which appears in
Eq.~(\ref{Eq:discretize}),
corresponds to the velocity $v\sub{g}$ in Eq.~(\ref{Eq:envelope}).
Even though the dimension of $\nu\sub{d}$
is different from that of velocities,
we hereafter call it ``velocity.''
It is the number of stages passed by the pulse 
per unit time.

\subsection{Spectrum condition}

The above discretization is valid
when the envelope $\E(x, t)$ is smooth enough that
the changes of $\E(x, t)$ over $\Delta x$ can be neglected.
From Eq.~(\ref{Eq:discretize}),
the fractional change between $v_n(t)$ and $v_{n+1}(t)$ is estimated as
\begin{align}
 \left|\frac{v_{n+1}(t)-v_n(t)}{v_n(t)}\right| \simeq 
 \frac{1}{\nu\sub{d}} \left|  \frac{1}{v_n(t)} \frac{d v_n(t)}{dt} \right|
 < \frac{\delta\omega}{\nu\sub{d}}
\end{align}
where we have assumed $v_n(t) \simeq v_{n+1}(t)$
and $\delta\omega$ is the bandwidth of $v_n(t)$.
If the condition,
\begin{align}
 \delta\omega / \nu\sub{d} = \delta\omega\cdot T \ll 1, \label{Eq:condition}
\end{align}
is satisfied, the discretization is justified.
This condition, $\delta \omega \cdot T \ll 1$
is accidentally coincide with the condition
that each all-pass filter works as an ideal group-delay
circuit; $H(\omega)\sim \exp(-i\omega T)$.
Thus we have found that the spectrum of the input signals should be restricted
within $\delta\omega \ll \nu\sub{d}$.

\section{Experiment\label{Sec:experiment}}

\subsection{Circuit parameters and input signal}
\begin{figure}[tb]
 \begin{center}
  \psfrag{time}[c][c]{$t$ / s}
  \psfrag{n}[c][c]{$n$}
  \psfrag{voltage}[c][t]{$v_n(t) / {\rm V}$}
  \includegraphics[]{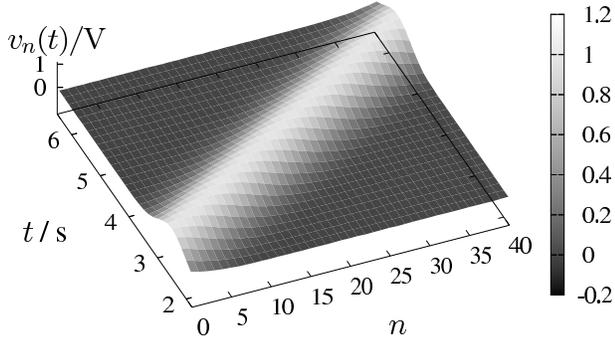}
  \caption{Simulation of wave propagation at constant speed
  $v\sub{d}=13{\rm /s}$.}
  \label{Fig:normal}
 \end{center}
\end{figure}

\begin{table}[]
\begin{tabular}{ll|l|l|l}
 \hline
 SW1  & SW2    & $R$  & $T (=2RC)$ & $\nu\sub{d} (=1/T)$ \\
 \hline \hline
 ON  & ON  
 & $476 \,{\rm k\Omega}$ & $0.078 \,{\rm s}$
 & $13 \,{\rm /s}$\\
 ON  & OFF & $909 \,{\rm k\Omega}$ & $0.15  \,{\rm s}$
 & $6.7 \,{\rm /s}$\\
 OFF & OFF & $10  \,{\rm M\Omega}$ & $1.6   \,{\rm s}$
 & $0.62 \,{\rm /s}$\\
 \hline
\end{tabular}
 \caption{Switch configurations and parameters.
$C=82\,\U{nF}$.}
 \label{Table:delay}
\end{table}

The experimental setup is shown
in Fig.~\ref{Fig:allpass-p} and 
the parameters of the delay circuit are shown in the inset.
Forty delay circuits are cascaded.
Each of the output is monitored with an LED.
The delay time $T=2RC$, or the speed $\nu\sub{d}=1/T$,
can be changed electronically with two analog switches (DG441)
connected to the resistors.
We show the relations between the switch configurations
and the constants ($R$, $T$, and $\nu\sub{d}$) in Table \ref{Table:delay}
\@.

The arbitrary waveform generator (Tektronix: AFG320)
provides Gaussian-like pulses
to the input of the delay circuits.
The waveform is given by
\begin{align}
 v_0 (t) = 
\begin{cases}
 0 & (t<0, \quad t>5\,\U{s}),\\
 V_0 \, e^{-4 (t-t\sub{s})^2/(\delta \tau)^2} & (0 \leq t \leq 5\,\U{s}),
\end{cases}
\end{align}
where $V_0=1.0\,\U{V}$, $t\sub{s}=2.5\,\U{s}$, and 
$\delta \tau=1.0\,\U{s}$.
The frequency spectrum of $v_0(t)$ is limited within
$\delta\omega \sim 4/\delta \tau=4.0 \,\U{Hz}$.

\subsection{Constant speed propagation}

Figure \ref{Fig:normal} shows the experimental result
for $\nu\sub{d}=13\,\U{/s}$.
The Gaussian-shaped pulse propagates down at the speed of 
13 circuits per second with its shape unchanged.
In this case,
the spectrum condition,
Eq.~(\ref{Eq:condition})\@, is satisfied
as $\delta\omega / \nu\sub{d} \sim 0.31 < 1$,
which guarantees that the pulse travels with small distortion.
The pulse length is estimated to be $\delta n=\nu\sub{d} 
\delta \tau  \sim 13$.

\subsection{Pulse length compression
\label{Sec:compression}}

\begin{figure}[]
 \begin{center}
  \psfrag{time}[c][c]{$t$ / s}
  \psfrag{n}[c][c]{$n$}
  \psfrag{voltage}[c][t]{$v_n(t) / {\rm V}$}
  \psfrag{tau}[c][c]{$\delta \tau$}
  \psfrag{n1}[c][c]{$\delta n$}
  \psfrag{n2}[c][c]{$\delta n$}
  \includegraphics[]{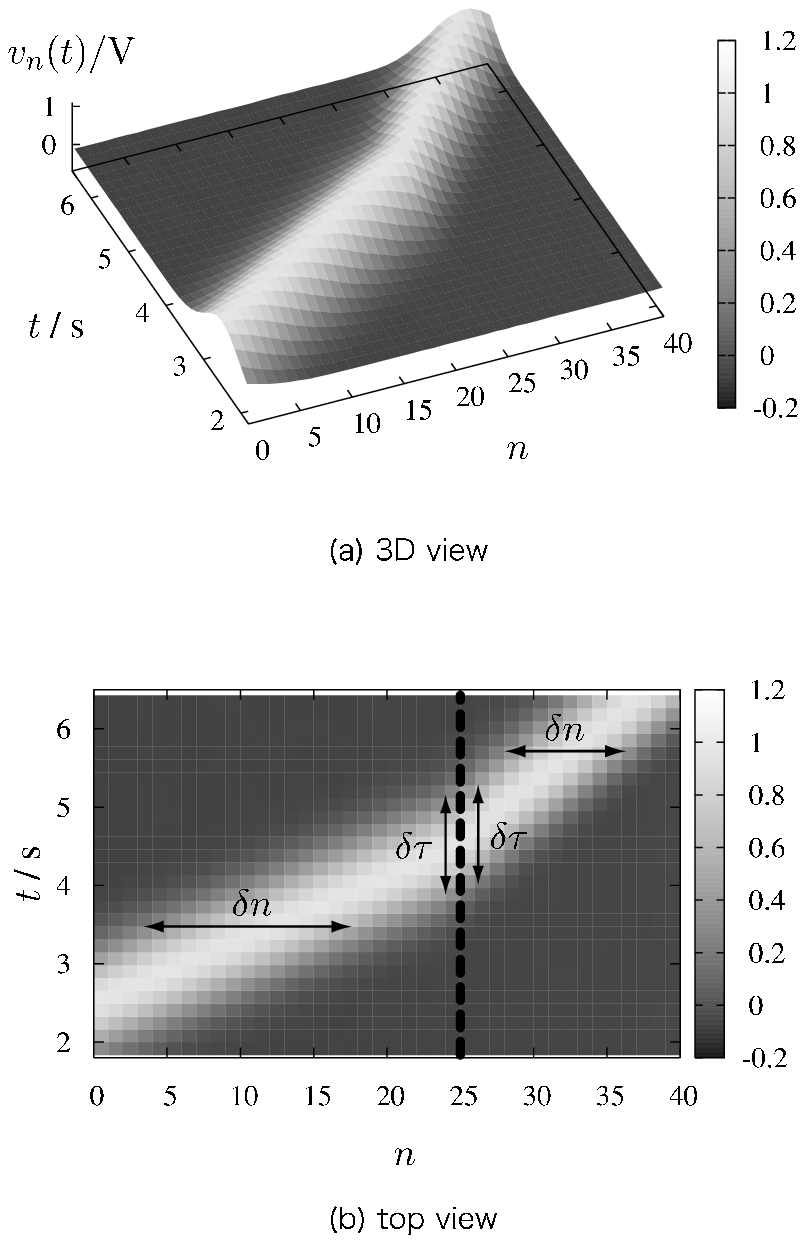}
  \caption{Pulse length compression.
  Pulse which runs at $v_1=13{\rm /s}$ in $n<25$,
  slows down to $v_2=6.7{\rm /s}$ in $n \geq 25$. }
  \label{Fig:compress1}
\end{center}
\end{figure}

\begin{figure}[]
 \begin{center}
  \psfrag{time}[c][c]{$t$ / s}
  \psfrag{n}[c][c]{$n$}
  \psfrag{voltage}[c][t]{$v_n(t) / {\rm V}$}
  \includegraphics[scale=1]{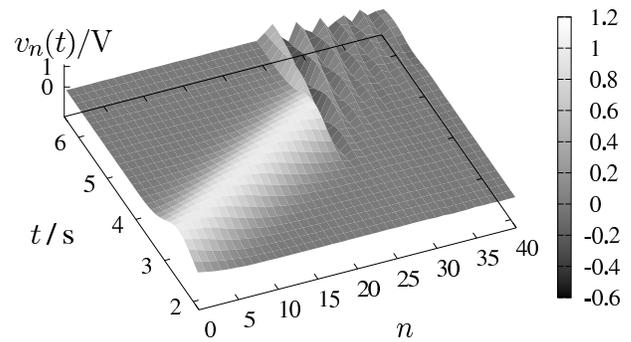}
  \caption{Experimental result for $v_1=13{\rm /s}$ in $n<25$
  and $v_2=0.62{\rm /s}$ in $n \geq 25$. 
  The pulse is collapsed  in $n \geq 25$,
  because the spectrum condition is violated.}
  \label{Fig:compress2}
 \end{center}
\end{figure}

In this section,
we consider the situation
where the speed $\nu\sub{d}$ is dependent on the location $n$.
We divide the cascaded circuits into two regions;
$n < 25$ (Region I) and $n \geq 25$ (Region II).
The velocity in each region is set as
\begin{align}
\nu\sub{d}(n)=
\begin{cases}
\nu_1 & (\mbox{Region I};\quad n< 25)\\
\nu_2 & (\mbox{Region II};\quad 25\leq n)
\end{cases}
.
\end{align}
This case corresponds to the light propagation experiment
where the pulse is fed from the vacuum (Region I)
into the EIT medium (Region II).

Figure \ref{Fig:compress1} 
shows the result for the case of $\nu_1=13\,\U{/s}$ and $\nu_2=6.7\,\U{/s} 
\, (\sim \nu_1/2)$.
At the boundary $n=25$, the pulse propagation is decelerated 
to almost half the speed.
It should be noted that 
the pulse has to be connected continuously at the boundary,
which is marked by the dashed line,
because the output of the 24th circuit is connected to the input of 
the 25th circuit.
Hence the pulse width $\delta \tau$ in time domain
is kept unchanged for any location $n$,
and the spectrum width $\delta\omega$ is also invariable.
In the present case, for both of the regions,
the spectrum condition is satisfied as
$\delta\omega / \nu_1 =0.31 < 1$, $\delta\omega / \nu_2 =0.60 <1$,
and then the Gaussian shape is almost maintained through the propagation.
On the other hand,
the pulse length $\delta n$ in Region II is compressed
into about half the length; $\delta n_1 = \nu_1 \delta \tau  \sim 13$
in Region I and $\delta n_2 =  \nu_2 \delta \tau \sim 6.7$
in Region II.
This is because the rear part of the pulse travels faster in Region I 
than the leading part runs in Region II
after the leading part reaches the boundary.
As mentioned before, the same pulse compression occurs
when the light pulse prepared in the vacuum
is fed into the EIT medium.

Next, we use the parameters: $\nu_1=13\,\U{/s}$, $\nu_2=0.62\,\U{/s}
\,(\sim \nu_1/20)$.
One might expect the pulse to be decelerated to $1/20$
and compressed spatially down to $1/20$.
But we obtain the result as shown in Fig.~\ref{Fig:compress2}.
In this case, the shape of the pulse is collapsed completely
in Region II.
This is because the spectrum condition is violated in Region II
($\delta\omega / \nu_2 = 6.4 > 1$).
The spectrum condition sets the lower speed limit:
$\delta\omega=4.0\,\U{/s}$.

In the experiment of the slow light by means of EIT,
similar spectrum condition exists.
If the spectrum of the light pulse is wider than the transparency
window of the EIT medium, the light is absorbed and 
the shape is strongly modified.

\subsection{Pulse freezing\label{Sec:freeze}}

Here, we deal with cases where the speed $\nu\sub{d}$ is a function of
time.
For simplicity, we divide the time interval into three periods and
in each period the velocity is kept constant;
\begin{align}
\nu\sub{d}(t)=
\begin{cases}
\nu_1 & (\mbox{Period I};\quad t< 4\,\U{s})\\
\nu_2 & (\mbox{Period II};\quad 4\,\U{s}\leq t < 7\,\U{s})\\
\nu_3 & (\mbox{Period III};\quad 7\,\U{s}\leq t)
\end{cases}
.
\end{align}

\begin{figure}[]
 \begin{center}
  \psfrag{time}[c][c]{$t$ / s}
  \psfrag{n}[c][c]{$n$}
  \psfrag{voltage}[c][t]{$v_n(t) / {\rm V}$}
  \psfrag{t1}[c][c]{$\delta \tau$}
  \psfrag{t2}[c][c]{$\delta \tau$}
  \psfrag{dn}[c][c]{$\delta n$}
  \includegraphics[]{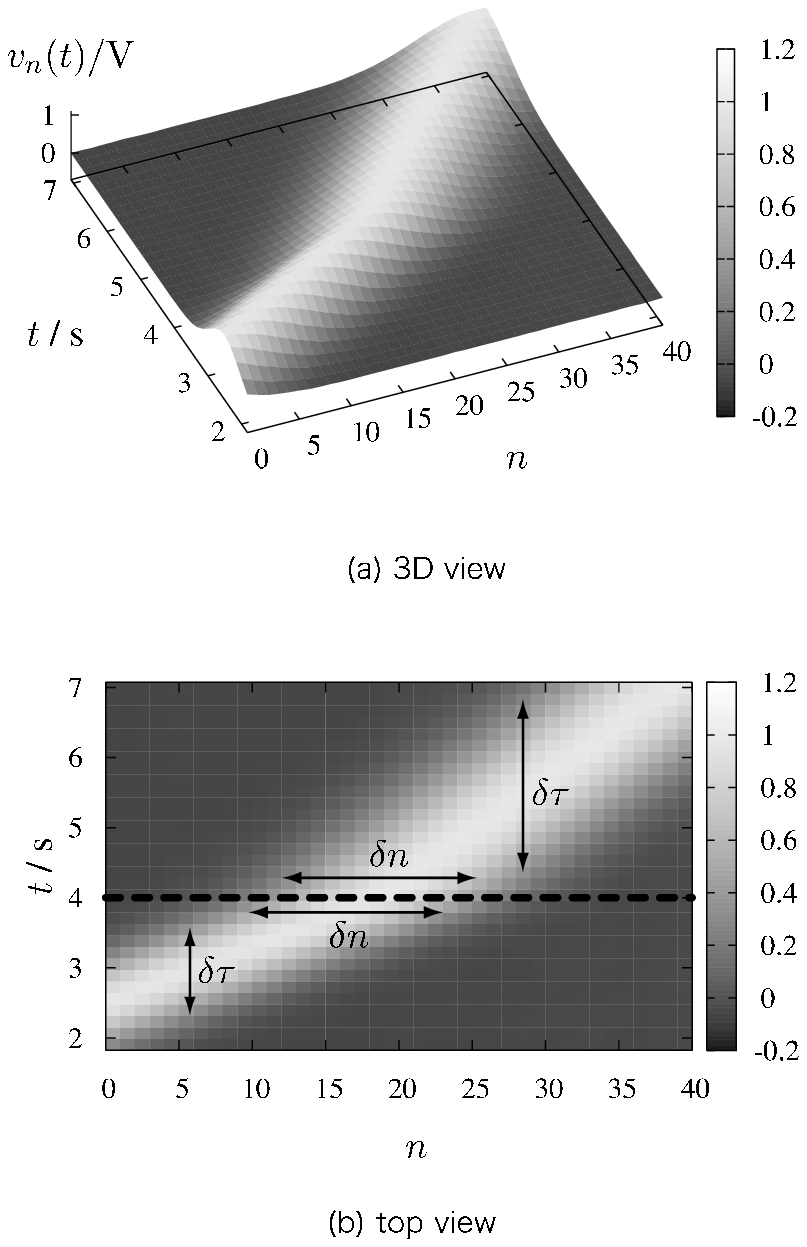}
  \caption{Changing the speed $v\sub{g}$ in time.
  Pulse which initially travels at $v_1=13{\rm /s}$
  slows down to $v_2=6.7{\rm /s}$}
  \label{Fig:spectrum_comp}
 \end{center}
\end{figure}

\begin{figure}[h]
 \begin{center}
  \psfrag{time}[c][c]{$t$ / s}
  \psfrag{n}[c][c]{$n$}
  \psfrag{voltage}[c][t]{$v_n(t) / {\rm V}$}
  \psfrag{t1}[c][c]{$\delta \tau$}
  \psfrag{t2}[c][c]{$\delta \tau$}
  \psfrag{dn}[c][c]{$\delta n$}
  \includegraphics[scale=1]{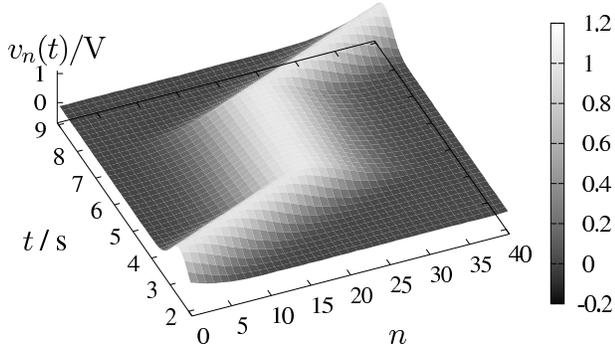}
  \caption{Pulse freezing.}
  \label{Fig:Freeze}
 \end{center}
\end{figure}

First we consider the case
$\nu_1=13\,\U{/s}$
and $\nu_2=\nu_3=6.7\,\U{/s}$.
The result is shown in Fig.~\ref{Fig:spectrum_comp}.
At the end of Period I ($t=4\,\U{s}$), 
the most of the pulse enters into the circuits.
The pulse is then slowed down from the leading edge to the 
rear edge instantaneously.
The pulse length $\delta n$ is unchanged,
but the pulse width in time domain, $\delta \tau$,
doubles after the boundary.
In other words, the frequency spectrum 
is compressed into half and $\delta\omega$ is halved.
This means that the $\delta\omega / \nu\sub{d}$ 
remains constant ($=0.31$).
This situation corresponds to the case
where we change the group velocity 
while the entire pulse is in the EIT medium ($0<n\leq 40$).

The violation of the spectrum condition, as discussed in the
previous section, 
can be avoided by the strategy of changing $\nu\sub{d}$ in time.
It is possible to attain much slower propagation or 
even to stop the pulse.
We modify the time sequence as
$\nu_1=13\,\U{/s}$,
$\nu_2=0.62\,\U{/s}$, and $\nu_3=13\,\U{/s}$.
The result is shown in Fig.~\ref{Fig:Freeze}.
In Period II,
the propagation of pulse is almost freezed,
and released at $t=7\,\U{s}$.
As mentioned above, the spectrum condition is always satisfied
because $\delta\omega / \nu\sub{d}$ is conserved in any time.
Hence, the pulse can propagate or be frozen without changing its shape
in the time-varying system.

In the EIT case, the change of the group velocity
is attributed to the dark state polariton.
The dark state polariton is a coupled mode of
the electric field and the atomic coherence.
The velocity of the polariton determined by the
mixing ratio, which depends on the control light intensity.
The atom-like polariton runs slower and the purely atomic
polariton stays still.
The pulse shape is imprinted in atomic states.
The storage time is limited by the relaxation time of
the atomic coherence.
In the circuit model with a resistor connected
parallel to each capacitor, we can simulate the decoherence.

\section{Summary\label{Sec:discuss}}

In this paper,
we introduced an electronic circuit which
simulates slow light and stopped light.
Most of the properties in the simulation are common to that of the
light propagation in the EIT medium.
However the mechanism of the velocity control is much more simple for the
case of simulating circuit.
Without knowing the details of EIT and dark state polaritons, we can understand
the physics of slow light and freezed light.
In addition,
the simulation with the circuits has the advantage
that the parameters can be changed easily.
This paper deals with three cases:
constant $v\sub{g}$,
position dependent velocity $v\sub{g}(x)$,
and time-varying velocity $v\sub{g}(t)$.
It will be interesting to consider the general case
where the velocity is a function of space and time,
$v\sub{g}(x, t)$ (for example \cite{Leonhardt}).
For cases where the
implementation with EIT methods is difficult,
the circuit model may likely be feasible.

The present model deals with envelopes and 
can describe only one-way wave propagations.
It cannot be used to reproduce bidirectional phenomena such as 
reflections and interference, where both the forward and backward 
waves come into play.
But it does reproduce one of the most fundamental properties of waves;
a disturbance propagates at a certain velocity keeping its shape
through a medium.

\section*{Acknowledgment}

We gratefully thank Vladan Vuletic for helpful discussion.
This research was supported by a Grant-in-Aid for Scientific Research
No.15654056 and No.~16740231,
and the 21st Century COE Program No.~14213201.

\appendix

\section{Polarization in a linear dispersive medium\label{appendix}}

In this appendix, we
derive the relation (\ref{Eq:polarization}) between the envelopes
$\E(t)$ and $\P(t)$.
For brevity, we omit the explicit dependence on $x$.
For linear media, the polarization $P(t)$ induced by the
electric field $E(t)$ is given by
\begin{align}
P(t)=\epsilon_0\intinf\hat{\chi}(t-\tau)E(\tau) d\tau,
\label{eq:a1}
\end{align}
where $\hat{\chi}(t)$ is the response function.
The dielectric susceptibility $\chi(\omega)$ and the
response function $\hat{\chi}(t)$ are connected with the
Fourier transform:
\begin{align}
\hat{\chi}(t)=\frac{1}{2\pi}
\intinf\chi(\omega)e^{i\omega t} d\omega.
\end{align}
The realness of $\hat{\chi}(t)$ implies $\chi(-\omega)=\chi^*(\omega)$.
Equation (\ref{eq:a1}) can be Fourier-transformed as
\begin{align}
\hat{P}(\omega)=\epsilon_0\chi(\omega)\hat{E}(\omega)
.
\label{eq:a3} 
\end{align}
The Fourier transform of $E(t)=\E(t) \, e^{i \omega_0 t}+\cc$
can be expressed as,
\begin{align}
 \hat{E}(\omega) &= \intinf \E(t) \, e^{i \omega_0 t} e^{- i \omega t}
 d t\ + \intinf \E^*(t) \, e^{-i \omega_0 t} e^{- i \omega t} d t
 \nonumber \\
 &= \hat{\E}(\omega-\omega_0) + \hat{\E}^*(-\omega-\omega_0),
 \label{eq:a4}
\end{align}
in terms of the Fourier transform of the envelope:
\begin{align}
 \hat{\E} (\Omega) = \intinf \E(t) \, e^{-i \Omega t} dt.
\end{align}
Assuming that the spectrum of the electric field is 
restricted around $\pm \omega_0$,
we can use the Taylor expansion of the susceptibility $\chi(\omega)$:
\begin{align}
 \chi(\omega) \simeq \chi_0 + \chi_1 ( |\omega|-\omega_0),
\label{eq:a6}
\end{align}
where $ \chi_0  \equiv \chi(\omega_0)$,
$\chi_1 \equiv d\chi/d\omega (\omega_0)$.
When the losses in the medium are negligible, 
the susceptibility is real and must be symmetric;
$\chi(-\omega)=\chi(\omega)$.
The moduli has been introduced in order to take into account
the symmetry.

From Eqs.~(\ref{eq:a3}), (\ref{eq:a4}) and (\ref{eq:a6}), we have
\begin{align}
 \hat{P}(\omega) 
 &\simeq \epsilon_0 \chi_0 \hat{\E}(\omega-\omega_0) 
 + \epsilon_0 \chi_1 (\omega-\omega_0) \hat{\E}(\omega-\omega_0) 
 \nonumber \\
 & \quad {} + \epsilon_0  \chi_0 \hat{\E}^*(-\omega-\omega_0)
 + \epsilon_0 \chi_1 (-\omega-\omega_0) \hat{\E}^*(-\omega-\omega_0)
.
\label{app1}
\end{align}
and its inverse Fourier transform yields
\begin{align}
 P(t) &= \epsilon_0 \left(
 \chi_0 \E(t) \, e^{i \omega_0 t} - i \chi_1 \frac{d \E}{d t}
 \, e^{i \omega_0 t} \right) + \cc \nonumber \\
 &= \P (t) e^{i \omega_0 t} + \cc,
\end{align}
where we have used,
\begin{align}
 \E(t)&=\frac{1}{2\pi}\intinf\hat{\E}(\Omega)e^{i\Omega t}d\Omega,
 \nonumber \\
\frac{d}{d t} \E(t)
&=\frac{1}{2\pi}\intinf(i\Omega)\hat{\E}(\Omega)e^{i\Omega t}d\Omega
.
\end{align}
Now we have
Eq.~(\ref{Eq:polarization}):
\begin{align}
 \P(t) = \epsilon_0 \left(
 \chi_0 - i \chi_1 \frac{d}{d t} \right) \E(t).
\end{align}
It should be noted that only the second term contributes 
to the group velocity in deriving of Eq.~(\ref{Eq:envelope}).

\end{document}